\documentclass[fleqn,usenatbib]{mnras}

\usepackage{newtxtext,newtxmath}

\usepackage[T1]{fontenc}
\usepackage{ae,aecompl}

\usepackage{lineno}
\usepackage{gensymb}
\usepackage{graphicx}	
\usepackage{amsmath}	
\usepackage{amssymb}	

\newcommand\omegax{\Omega_{\rm X0}}
\newcommand\omegam{\Omega_{\rm m0}}
\newcommand\omegar{\Omega_{\rm r0}}

\newcommand{\kB}{k_{\rm B}}

\newcommand{\todo}[1]{{\tt TODO: #1}}

\title[Hot matter analytic solutions]{Too hot to handle? Analytic solutions for massive neutrino or warm dark matter cosmologies}

\author[Slepian \& Portillo]{
Zachary Slepian$^{1,2}$\thanks{E-mail: zslepian@lbl.gov (ZS)} \&
Stephen K. N. Portillo$^{3}$\thanks{E-mail: sportillo@cfa.harvard.edu (SKNP)}
\\
$^{1}$Einstein Fellow, Lawrence Berkeley National Laboratory, 1 Cyclotron Road, Berkeley CA 94720\\
$^{2}${Berkeley Center for Cosmological Physics, University of California, Berkeley, CA 94720}\\
$^{3}$Harvard-Smithsonian Center for Astrophysics, 60 Garden Street, Cambridge, MA 02138
}

\date{Accepted XXX. Received YYY; in original form ZZZ}

\pubyear{2017}

\begin{document}
\label{firstpage}
\pagerange{\pageref{firstpage}--\pageref{lastpage}}
\maketitle

\begin{abstract}
We obtain novel closed form solutions to the Friedmann equation for cosmological models containing a component whose equation of state is that of radiation $(w=1/3)$ at early times and that of cold pressureless matter $(w=0)$ at late times. The equation of state smoothly transitions from the early to late-time behavior and exactly describes the evolution of a species with a Dirac Delta function distribution in momentum magnitudes $|\vec{p}_0|$ (i.e. all particles have the same $|\vec{p}_0|$). Such a component, here termed ``hot matter'', is an approximate model for both neutrinos and warm dark matter. We consider it alone and in combination with cold matter and with radiation, also obtaining closed-form solutions for the growth of super-horizon perturbations in each case. The idealized model recovers $t(a)$ to better than $1.5\%$ accuracy for all $a$ relative to a Fermi-Dirac distribution (as describes neutrinos). We conclude by adding the second moment of the distribution to our exact solution and then generalizing to include  all moments of an arbitrary momentum distribution in a closed form solution.
\end{abstract}



\section{Introduction}
The effects of massive neutrinos on structure formation have been of considerable recent interest in cosmology (\citealt{Lesgourguesrvw} for a review of the theory). We are now likely on the cusp of experiments that will determine the mass of the heaviest one or two neutrinos and also whether the mass hierarchy is normal or inverted (see \citealt{Abazajian15} for a review). Massive neutrinos have an unusual cosmological behavior: unlike any other components, we know with certainty that they have a time-varying equation of state, which behaves like that of radiation at high redshift but like that of matter nearer the present. As a consequence neutrino perturbations do not grow until late times, when the neutrinos behave like matter, and this suppresses the overall amplitude of matter perturbations observed at present on scales smaller than roughly the neutrino horizon at this transition point. \citet{Bond80} solve for the growth on scales much smaller than the free-streaming scale, and \citet{HuEisenstein98}, \citet{Eisenstein99} obtained fitting formulae anchored by this small-scale solution and the large-scale behavior, which is simply the linear growth rate. The suppression in the matter power spectrum (e.g. \citealt{Hu98}, \citealt{Takada06}, \citealt{Bird12}) and bispectrum (\citealt{Levi16}) will be used to probe the neutrino masses with upcoming large-scale redshift surveys such as DESI \citep{Levi13} and Euclid \citep{Laureijs11}. The neutrino mass also impacts both the linear growth rate of perturbations and creates scale-dependent bias, effects which given multiple tracers and a high enough number density can enable measurements below cosmic variance \citep{LoVerde16}. The CMB lensing power spectrum and bispectrum can also be used to constrain the neutrino masses (\citealt{Namikawa10}, \citealt{Wu14}, \citealt{Abazajian15}, \citealt{Allison15}, \citealt{Namikawa16}).

A second, more speculative case where a time varying equation of state transitioning from radiation-like to matter-like is relevant is warm dark matter (WDM), a modification to the standard cold dark matter (CDM) paradigm that has the DM behave as a relativistic species at very high redshift (for a review of DM candidates including WDM, see \citealt{Feng10}). For any massive particle, this behavior generically occurs when the Universe is hot enough that the particle's kinetic energy is of the order of its rest mass energy, so even standard CDM would have had this behavior at sufficiently early times. However for a WIMP with mass $\sim\!100\;{\rm GeV}$, this would occur at $z\sim\!3\times 10^{14}$, well before weak interactions went out of equilibrium meaning that any signature of a relativistic-non-relativistic transition would be washed out. Further, at these very high redshifts, after the transition the DM quickly becomes negligible compared to the radiation. WDM models thus generally focus on much lighter particles whose transition from relativistic to non-relativistic would occur late enough to have observable effects. We point out that if the WDM is produced in a resonant process in the early universe such that all particles share the same momentum magnitude, then the Delta-function distribution function toy model studied in the bulk of this paper is exact. We also note that the current work applies to models where the dark matter interacts with a bath of dark radiation that supplies pressure, such as recently discussed in \cite{Buckley14}, \cite{Foot15}, and \cite{CyrRacine16}.

In this work, we analyze a toy model that can describe both of these cases. We consider a massive species whose distribution of momentum magnitudes is a Dirac Delta function centered on some momentum of present-day magnitude $p_0$, and derive the exact equation of state that this component satisfies as a function of time. At early times, this component behaves like a massless particle, with equation of state $w = 1/3$, while at late times the component behaves like cold, pressureless matter, with $w=0$. The transition point occurs when the Universe's temperature is of order the particle's mass. We then integrate this equation of state to obtain the evolution of the density in this component. This form for the density can then be inserted into the Friedmann equation for evolution of the scale factor, and the Friedmann equation solved by quadrature. Finally, we show how to generalize our toy model to include higher moments of the distribution function, first showing how to add the second moment for e.g. a Fermi-Dirac distribution, as applies to neutrinos, and then showing how to add an arbitrary number of moments as long as an expansion of the distribution function in terms of its moments converges.  

To remind the reader that our solution is not specific to neutrinos or warm dark matter, we simply call this component ``hot matter.'' However we do focus on cases motivated by these two models: we solve matter plus hot matter, the relevant case for neutrinos, and radiation plus hot matter, relevant for warm dark matter. For the former, one might ideally solve matter plus hot matter plus radiation, as the radiation energy density is still dynamically important when the neutrinos become non-relativistic, even though matter is dominant. However we were unable to find closed form solutions in this case.

Much work has been done on theoretical modeling and observational constraints for both the neutrino mass and warm dark matter models. Recent cosmological constraints on the sum of the neutrino masses are around 15-20 meV (\citealt{Cuesta16}, \citealt{Giusarma16}, \citealt{Huang16}, \citealt{Vagnozzi17}), and combining cosmological and laboratory probes slightly tightens this bound (\cite{Gerbino16}). There are a variety of methods of constraining warm dark matter models; the current mass lower bound is $\sim\!3.3$ keV from the Lyman-$\alpha$ forest (\citealt{Viel13}). 

We now briefly detail previous work on closed-form solutions of the Friedmann equations. Single-component models (radiation, matter, curvature, cosmological constant) have well-known solutions both for the background and the growth of perturbations (see e.g. \citealt{Peebles}, \citealt{Padmanabhan}, \citealt{Ryden}). Several two-component models also have solutions, such as cosmological constant plus matter \citep{Weinberg} and radiation plus matter \citep{Ryden}.

\cite{Edwards72} solves the Friedmann equation containing matter, curvature, and a cosmological constant, while \cite{Edwards73} does so for a model with radiation, matter, curvature, and a cosmological constant.

\cite{Chen_cheby} treats multi-component models with a cosmological constant and in arbitrary spatial dimension using Chebyshev's theorem to analyze the integrability; this theorem describes whether integrals that typically appear in the solutions can be carried out in closed form.
\cite{Chen_nonlin} applies Chebyshev's theorem to single-component models with non-linear equations of state. \cite{Chen_roulette} uses roulettes---curves generated by tracing the path of a fixed point on a closed shape as it is rolled along a line (e.g. a circle gives rise to the cycloid)---to further explore analytic solutions of the Friedmann equation beyond the regime probed by Chebyshev's theorem.

The growth of perturbations has also been solved for some two-component models (matter and radiation: \citealt{Groth75}, \citealt{Kodama84}).
\cite{EdwardsHeath76} solves for the growth of perturbations in models with matter, curvature, and cosmological constant, while \cite{Heath77} considers whether these perturbations become bound. \cite{Dem05} analytically treats the growth of perturbations in models with all combinations of the standard cosmological parameters (cosmological constant, curvature, matter, and radiation), providing a useful set of references to previous work solving cases as well as adding new analysis. In some cases there is no closed-form solution available but only series around singular points. In particular, the most general model has a growth equation of Fuchs-type, with six singular points, and solutions can be written in series around these points. The growth of matter perturbations seeded by ``causal'' sources such as topological defects from symmetry-breaking in the early Universe has also been investigated using a Green's function formalism \citep{ProtyWu00}.

There has also been considerable work on related integrals to those giving quadrature solutions of the Friedmann equation. In particular, many exact solutions exist for the distance-redshift relation for cosmologies with multiple components or non-standard equations of state. These are succinctly reviewed in \cite{Eisenstein97} and \cite{Baes17}. 

An additional application of the current work is where the standard dark matter is taken to have a small non-zero pressure due to the generation of peculiar velocities by the growth of structure. This scenario, known as ``backreaction'', has been invoked to explain accelerated expansion in lieu of dark energy, though it seems insufficient to do so (e.g. \citealt{Seljak96}, \citealt{Mukhanov97}, \citealt{Kaiser17}). However if one wished to account for the small DM pressure due to ``backreaction,'' the formalism of this paper enables so doing. Dark matter thermal velocity dispersions do have a small effect on the growth of structure, and the impact on CMB lensing and the matter power spectrum has been studied in e.g. \cite{Piattella16}.

We highlight that the primary advance of the current work is to capture the evolution of a component with equation of state that follows a smooth, physical trajectory in time from hot to cold, as well as to outline a method for incorporating the effects on the cosmological expansion rate of an arbitrary distribution function for the species. We solve a number of cases: hot matter only (\S\ref{sec:single_cmpt}), hot matter plus radiation (\S\ref{sec:hm_r}), and hot matter plus matter (\S\ref{sec:hm_m}). We also analyze the growth of super-horizon perturbations (i.e. where only gravity affects the evolution and pressure perturbations are negligible) in each of these models, using a technique developed in \cite{SEBAO}; this discussion is in respectively \S\ref{sec:sh_perts_one_cmpt}, \S\ref{sec:hm_r_perts}, and \S\ref{sec:hm_m_perts}. Finally, in \S\ref{sec:approx_conv} we generalize our treatment to any momentum distribution that can be expanded in terms of its moments.

\section{Evolution of the hot matter density}
\label{sec:dens_evoln}
In the framework of General Relativity (GR), the Universe's cosmological expansion is related to its contents' energy densities by the Friedmann equation. The time-evolution of these energy densities is in turn determined by their equation of state, the ratio of pressure $P$ to energy density $\rho$. The relationship between time-evolution and equation of state occurs because for adiabatic expansion, the $p dV$ work done must be balanced by a reduction in energy: the equation of state determines the price paid in pressure for a given change in energy.

Consequently, to determine the Universe's evolution, two different equations must be solved: one for the background density's evolution as a function of scale factor, and a second for the scale factor as a function of time.

In this section we solve for the evolution of the hot matter density as a function of scale factor, and use this result in the rest of the paper to then solve different cases for the evolution of the scale factor as a function of time.

Assuming the distribution of hot matter momenta is a Dirac delta function centered at $|\vec{p}|=p$, the hot matter pressure is
\begin{align}
P=\frac{1}{3}npv=\frac{1}{3}\frac{npc}{\sqrt{(mc^2/pc)^2+1}},
\end{align}
where $n$ is the number density of hot matter particles, $m$ is their mass, and $c$ is the speed of light. The hot matter energy density is
\begin{align}
\rho=n\sqrt{(mc^2)^2 + (pc)^2}=npc\sqrt{(mc^2/pc)^2+1}.
\end{align}
The equation of state is then
\begin{align}
w=\frac{P}{\rho}=\frac{1}{3}\frac{1}{(mc^2/pc)^2+1}.
\end{align}
If the hot matter momenta have a Dirac delta function distribution at one time and are modified only by the Universe's expansion, then the hot momenta will have a Delta function distribution at future times centred at $|\vec{p}|=p_0/a$, where $p_0$ is the centre of the momentum distribution at $a=1$ and $a$ is the scale factor. Thus, the hot matter equation of state becomes
\begin{align}
w(a)=\frac{1}{3}\frac{1}{(mc^2a/p_0c)^2+1}=\frac{1}{3}\frac{1}{f_0^{-2}a^2+1}
\end{align}
where $f_0\equiv p_0c/(mc^2)$ is the  ratio of kinetic to rest mass energy at present, when we have set $a=1$. We note that the species' transition from relativistic to non-relativistic occurs when $a=f_0$. 

Integrating the continuity equation shows that
\begin{align}
\rho(a) &= \rho_0\exp\bigg[-3\int_1^{a}\frac{da'}{a'}[1+w(a')]\bigg]\nonumber\\
&=\rho_0 a^{-3}\exp[-3I_w(a)],
\end{align}
with 
\begin{align}
I_w(a)\equiv \frac{1}{3}\int_1^{a(t)}\frac{da'}{a'}w(a').
\end{align}
The required integral may be evaluated using the partial fractions decomposition 
\begin{align}
\frac{1}{a'}\frac{1}{f_0^{-2}a'^2+1} =\frac{1}{a'}-\frac{a'}{f_0^2+a'^2}
\end{align}
or the trigonometric substitution $f_0^{-2}a'^2=\tan^2\theta$.  We find
\begin{align}
I(a)=\frac{1}{6}\ln\left[\frac{1+f_0^2a^{-2}}{1+f_0^2} \right]
\end{align}
leading to 
\begin{align}
\rho(a)=\rho_0g^{1/2}(a)a^{-3}
\label{eqn:rho_of_a}
\end{align}
with
\begin{align}
g(a)=\frac{1+f_0^2a^{-2}}{1+f_0^2}.
\label{eqn:g_defn}
\end{align}

\section{Friedmann equation for hot-matter-only cosmology}
\label{sec:single_cmpt}
The Friedmann equation for this cosmology is
\begin{align}
H^2=\left(\frac{1}{a}\frac{da}{dt} \right)^2=H_0^2\omegax a^{-3}g^{1/2}(a),
\end{align}
where $H$ is the Hubble parameter, defined by the first equality, and $H_0$ is its value at present, when $a=1$. $\omegax$ is the hot matter density at present, and $g(a)$ is defined in equation (\ref{eqn:g_defn}). Taking the square root of both sides, rearranging, and integrating yields $t(a)$:
\begin{align}
H_0\sqrt{\omegax}t= \int_0^{a(t)}\frac{a^{1/2}da'}{g^{1/4}(a')} \equiv I_0(a).
\label{eqn:integral}
\end{align}
The required integral is
\begin{align}
I_0(a) = \frac{2}{3}(1+f_0^2)^{1/4}\left[-f_0^{3/2}+(f_0^2+a^2)^{3/4}\right].
\label{eqn:I0}
\end{align}
As $f_0\to 0$, we should have a matter-only cosmology; by inspection one can see that we recover
\begin{align}
a(t)=\left(\frac{t}{t_{\rm m0}}\right)^{2/3},\;\;\;t_{\rm m0}=\frac{2}{3H_0\sqrt{\omegam}}.
\label{eqn:matter_only}
\end{align}
As $f_0\to\infty$, we should have a radiation-only cosmology.  Note that we can make $f_0$ arbitrarily larger than any given scale factor $a$, and hence we can define $\epsilon =a/f_0\ll 1$.  Taylor-expanding $(f_0^2+a^2)^{3/4}\approx f_0^{3/2}(1+(3/4)\epsilon^2)$ and inserting this result into equation (\ref{eqn:I0}) yields
\begin{align}
\frac{2}{3}(1+f_0^2)^{1/4}\left[-f_0^{3/2}+f_0^{3/2}(1+\frac{3}{4}\epsilon^2) \right]=H_0\sqrt{\omegar}t.
\end{align}
Further noting that $f_0^2\gg1$ in the first factor, simplifying, and solving for $a(t)$ yields the usual radiation-dominated
\begin{align}
a(t)=\left(\frac{t}{t_{\rm r0}} \right)^{1/2},\;\;\;t_{\rm r0}=\frac{1}{2H_0\sqrt{\omegar}}.
\label{eqn:radiation_only}
\end{align}

We now solve for $a(t)$ in the general case explicitly; it is algebraic in $t$.  Using equations (\ref{eqn:integral}) and (\ref{eqn:I0}) we obtain
\begin{align}
a(t)=\left[\left(\frac{t}{t_{\rm fid}}+f_0^{3/2}\right)^{4/3}-f_0^2\right]^{1/2},\;\;\;t_{\rm fid}=\frac{2(1+f_0^2)^{1/4}}{3H_0\sqrt{\omegax}}.
\label{eqn:a_of_t}
\end{align}
$t_{\rm fid}$ is a fiducial time chosen to simplify $a(t)$ and is not the age of the Universe in this model, though it is the age of the Universe in the limit that $f_0 \to 0$ (cold matter). 

By solving for the time when $a=1$, we find the age of the Universe as
\begin{align}
t_0 = t_{\rm fid}\bigg[(1+f_0^2)^{3/4} -f_0^{3/2}\bigg].
\label{eqn:hm_only_age}
\end{align}
We note that the age of the Universe does depend on $\omegax$ through the dependence of $t_{\rm fid}$ on this quantity as in equation (\ref{eqn:a_of_t}).

The $f_0\to 0$, matter-only limit recovers equation (\ref{eqn:matter_only}) by inspection, while Taylor-expanding to leading order in $(t/t_0)/f_0^{3/2}$ recovers the radiation-only, $f_0\to\infty$ limit given by equation (\ref{eqn:radiation_only}). The age of the Universe remains finite in this limit because $t_{\rm fid}$ diverges exactly as the second factor in equation (\ref{eqn:hm_only_age}) goes to zero, and we recover the radiation-dominated result.

\begin{figure}
\includegraphics[width=1.1\columnwidth]{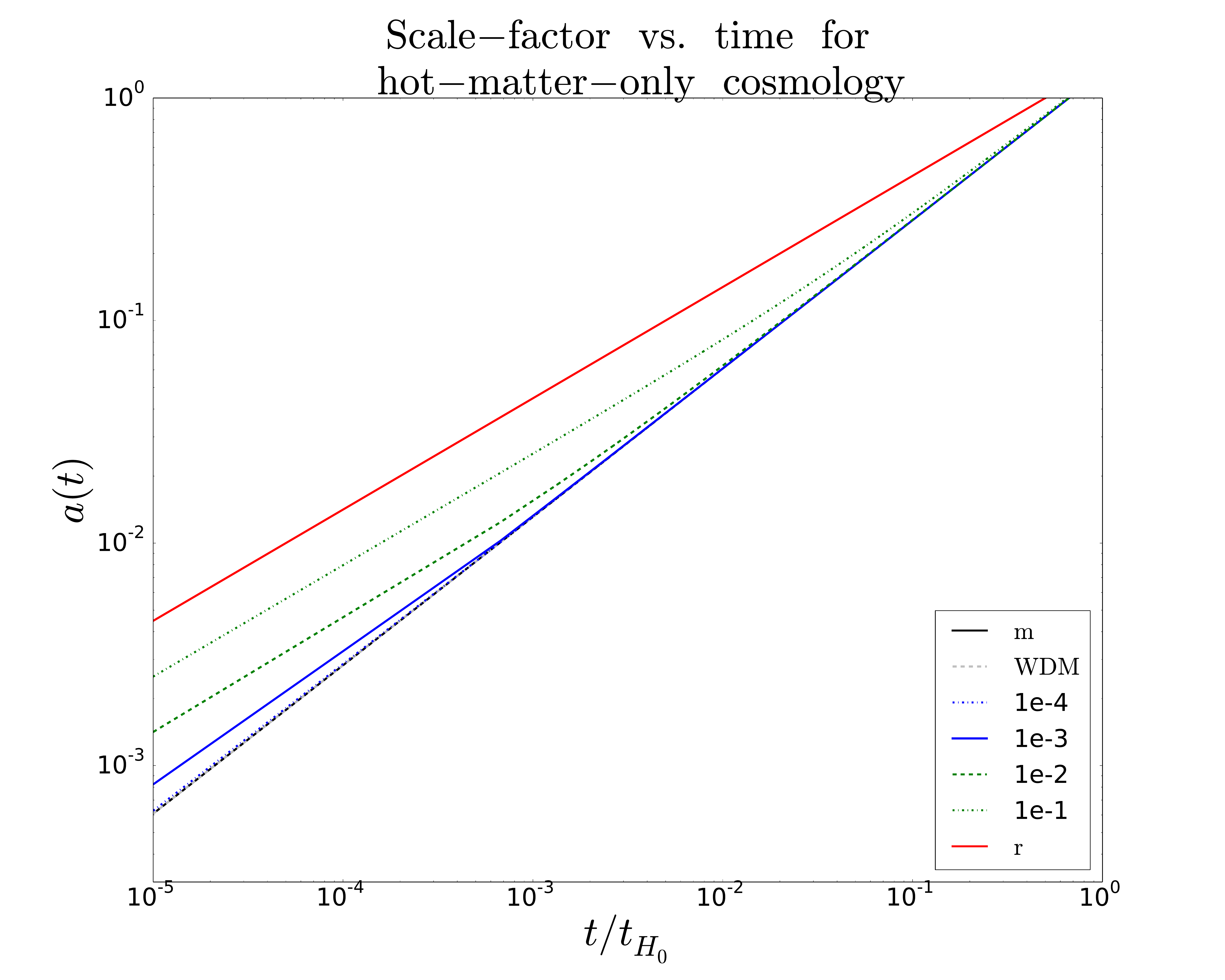}
\caption{Scale-factor vs. time for illustrative values of $f_0$ in the hot-matter-only cosmology, with the values given in the legend. ``${\rm m}$'' denotes matter-only $(f_0 = 0)$, and ``${\rm r}$'' denotes the radiation-only limit. Time is in units of the Hubble time $t_{H_0} = 1/H_0$ and $\Omega_{{\rm X}0} = 1$. The expansion is a power law $a \propto t^{1/2}$ for matter-only and $a \propto t^{2/3}$ for radiation-only, and the hot-matter solutions lie between these limits. When the ratio of kinetic energy to rest energy at a given scale-factor, $f(a) = f_0/a$, dips below unity, the hot matter has become non-relativistic, and its evolution becomes more similar to matter-dominated. Thus for the $f_0 = 0.1$ model the blue curve matches onto the black (matter-dominated) curve around $a \simeq 0.1$.}
\label{fig:hm_only}
\end{figure}

Our solution to the Friedmann equation in the hot-matter-only case is shown in Figure \ref{fig:hm_only}. In this Figure only, we choose a large range of values of $f_0$, the present-day ratio of kinetic to rest-mass energy in the hot matter. We show the matter-only limit ($f_0 = 0$) and the radiation-only limit ($f_0\to\infty$) as well as the value of $f_0$ corresponding to the current lower bound on the mass of Warm Dark Matter (WDM), which is of order $f_0\simeq 10^{-7}$. We also show four decades of $f_0$ between $10^{-4}$ and $10^{-1}$, all of which have become non-relativistic by the present ($f_0 = 1$ is the cut-off above which the hot matter would still be relativistic). 

Figure \ref{fig:hm_only} shows that the behavior of $a(t)$ smoothly moves from matter-like at small $f_0$ to radiation-like at large $f_0$. We are able to show a large range of values in this Figure because the effect of varying $f_0$ is most significant in a hot-matter-only model; in the other models we consider, which have additional components, we set $\omegax = 0.5$, and so it becomes more difficult to see differences in behavior as $f_0$ varies because these differences are diluted by the presence of the additional components. Therefore in the rest of this work (Figure \ref{fig:pert_hm_only} and onwards) we show a narrower range of values for $f_0$.

We note here a general trend that will become evident in all of the Figures to follow. The scale factor grows faster with time in a matter-dominated model than in a radiation-dominated model, and so those models with smaller $f_0$ will have lower amplitude in the past than those with larger $f_0$. This is because we have set $a=1$ at present; the lower-$f_0$ models grow faster and thus can reach $a=1$ at present from lower past values than can those with larger $f_0$.

However, perturbations to the dominant component grow faster in radiation-dominated models than in matter-dominated models. Thus the ordering of curves with $f_0$ reverses relative to the plots of scale factor; curves with lower $f_0$ have slower growth of perturbations and so must begin at higher amplitudes to reach $\beta =1$ at present, where we have normalized all cases to have perturbations equal to unity. 

This qualitative trend, that models with smaller $f_0$ have lower amplitude scale-factor in the past due to their faster growth of the scale-factor, but larger amplitude perturbations in the past due to their slower growth of perturbations, will recur in all the Figures in this work.

\section{Growth of super-horizon perturbations in the hot-matter-only cosmology}
\label{sec:sh_perts_one_cmpt}
We consider the growth of an overdense spherical perturbation (``bubble Universe'') of radius $r$ within an otherwise homogeneous background Universe.  The bubble will satisfy its own Friedmann equation, but it is now above the critical density and so will also have a curvature term.  The amplitude $C$ of the curvature at some fiducial time is set by the density perturbation at that time.  We have
\begin{align}
H_{\rm pert}^2=\left(\frac{\dot{r}}{r}\right)^2=H_0^2\left[g^{1/2}(r)r^{-3}+Cr^{-2} \right],
\end{align}
where we have chosen $H_{\rm pert}(r=1)=H_0(1+ C)$. $g(r)$ is as defined in equation (\ref{eqn:g_defn}) with $a$ replaced by $r$. Taking the square root of both sides, rearranging, multiplying numerator and denominator by $r'^{1/2}$, and integrating, we have the relation for the time that
\begin{align}
\int_0^{r}\frac{r'^{1/2}dr'}{\sqrt{g^{1/2}(r')+Cr'}}=H_0t.
\end{align}
The time measured in the perturbed, bubble universe and the time measured in the background universe are the same in synchronous gauge (see Slepian \& Eisenstein 2016, \S3 and \S4, for further discussion)---we term this the ``equal time'' condition. Our strategy will thus be to use this constraint to compute the value of $C$ in terms of a radial perturbation to the scale factor we introduce at some fiducial time. Physically, we are creating an overdense bubble universe by compressing the background universe slightly, so that its scale factor is reduced. This induces a curvature perturbation as the bubble universe will be closed rather than flat. Once we have related the curvature perturbation and the radial perturbation to the scale factor, we may then use the ``equal-time'' condition again to obtain the evolution of the radial perturbation.

Our first step is to Taylor-expand the integrand to leading order in $Cr'/g^{1/2}(r')$. Since $C$ is the curvature perturbation induced by the density perturbation, we have $Cr'/g^{1/2}(r')\ll 1$ for all $r'$. As $r'\to 0$, $g^{1/2}(r')\to \infty$, and $r'\to 0$ in the numerator $Cr'$ further guarantees the validity of our expansion in this limit. Physically, this is the statement that at the initial time any perturbation becomes vanishingly small. Meanwhile, as $r'\to \infty$, corresponding to late times $g^{1/2}(r')\to (1+f_0^2)^{-1}$ while $Cr'\to \infty$. Physically this is just the statement that in the linear regime, perturbations eventually become of order unity, at which point the perturbative expansion is no longer valid. We simply restrict to $r'$ sufficiently small that this case does not occur. This restriction corresponds to working at early enough times that linear perturbation theory is valid; recall $r'$ is the scale-factor of the perturbed bubble universe and is thus a proxy for time.

Consequently we may Taylor-expand the integrand in $C$ to find
\begin{align}
H_0t &= \int_0^{r} \frac{r'^{1/2}}{g^{1/4}(r')} -C\int_0^{r}  \frac{r'^{3/2}}{2g^{3/4}(r')}+\mathcal{O}(C^2)\nonumber\\
& =I_0(r) - CI_1(r) +\mathcal{O},
\label{eqn:I0_I1_defn}
\end{align}
where $I_0$ is simply our result from the unperturbed, background case of \S\ref{sec:single_cmpt}, defined in equation (\ref{eqn:I0}), while we can compute $I_1(r)$ explicitly as 
\begin{align}
&I_1(r)=\nonumber\\
&\frac{1}{5}(1+f_0^2)^{3/4}\left[4f_0^{5/2}-4f_0^2(f_0^2+r^2)^{1/4}+r^2(f_0^2+r^2)^{1/4}\right].
\label{eqn:I1}
\end{align}
 
We now obtain the evolution of a perturbation to the scale factor of the background universe. In particular, we take it that the scale factor of the bubble universe $r$ is related to that of the background universe $a$ as 
\begin{align}
r(a) = a(1-\beta(a)).
\end{align}
We note that the more usual density perturbation $\delta \equiv [\rho - \bar{\rho}]/\bar{\rho}$, with $\rho$ the perturbed density and $\bar{\rho}$ the background density, is related to $\beta$ as
\begin{align}
\delta(a) = 3\beta(a)
\label{eqn:delta_beta}
\end{align}
because taking a Taylor-series shows that perturbing the radius by $\beta$ alters the unperturbed volume $V_0$ as $V_0 \to V_0(1-3\beta)$, since $V\propto r^3$. This leads to a density enhancement $\bar{\rho} \to \bar{\rho}(1 + 3\beta)$ and using the definition of $\delta$ yields equation (\ref{eqn:delta_beta}).

We first compute the value of $C$ in terms of the perturbation to the scale factor $\beta_0$ at some fiducial time, which we take to be  $t_0$ (equivalently, $a=1$). We will then solve for the full time-evolution of $\beta$.

As discussed earlier, in synchronous gauge the background and perturbed bubble universe measure the same time. Thus
 \begin{align}
 H_0t=I_0(1) & = I_0(1-\beta_0)-CI_1(1-\beta_0)\nonumber\\
& \approx I_0(1)+\frac{dI_0}{dr}\bigg|_1\beta_0 -CI_1(1).
 \end{align}
In the second line we have taken Taylor series for $I_0$ and $I_1$ and retained only leading order terms (recall that both $C$ and $\beta_0$ are small, meaning we can drop the $\mathcal{O}(C\beta_0)$ contribution that would arise from $I_1$). Solving for $C$ yields
\begin{align}
C=\frac{1}{I_1(1)}\frac{dI_0}{dr}\bigg|_1\beta_0,
\label{eqn:C_soln}
\end{align}
where the required derivative is simply the integrand of equation (\ref{eqn:I0}).

We now find the time-evolution of the perturbation $\beta(a)$.The background Universe and perturbed Universe measure the same time not only at $t_0$, but at all times, so we further have that
\begin{align}
H_0t=I_0(a)&=I_0(a[1-\beta(a)])-CI_1(a[1-\beta(a)])\nonumber\\
&\approx I_0(a)-\frac{dI_0}{da}\bigg|_a a\beta(a) -CI_1(a).
\end{align}
Solving algebraically for $\beta(a)$ we obtain
\begin{align}
&\beta(a)=C\frac{I_1(a)}{a}\left( \frac{dI_0}{dr}\bigg|_a\right)^{-1}=\frac{I_1(a)}{I_1(1)}\frac{dI_0/dr|_1}{dI_0/dr|_a}\beta_0\nonumber\\
&=\frac{I_1(a)}{I_1(1)}\frac{g^{1/4}(a)}{a^{3/2}}\beta_0,
\label{eqn:perts_gen}
\end{align}
where we used equation (\ref{eqn:C_soln}) to obtain the second equality, which is general. For the third equality, we inserted equation (\ref{eqn:I0}) for $I_0$.  $I_1(a)$ is defined by equation (\ref{eqn:I1}) and $g(a)$ by equation (\ref{eqn:g_defn}).

Taking the limit as $f_0\to0$ should recover the growth of perturbations in a matter-only cosmology, i.e. $\beta(a)\propto a$.  From inspection of equation (\ref{eqn:I1})  in this limit $I_1(a)\to (1/5)a^{5/2}$, and using this result in equation (\ref{eqn:perts_gen}) we recover the desired scaling.

\begin{figure}
\includegraphics[width=1.1\columnwidth]{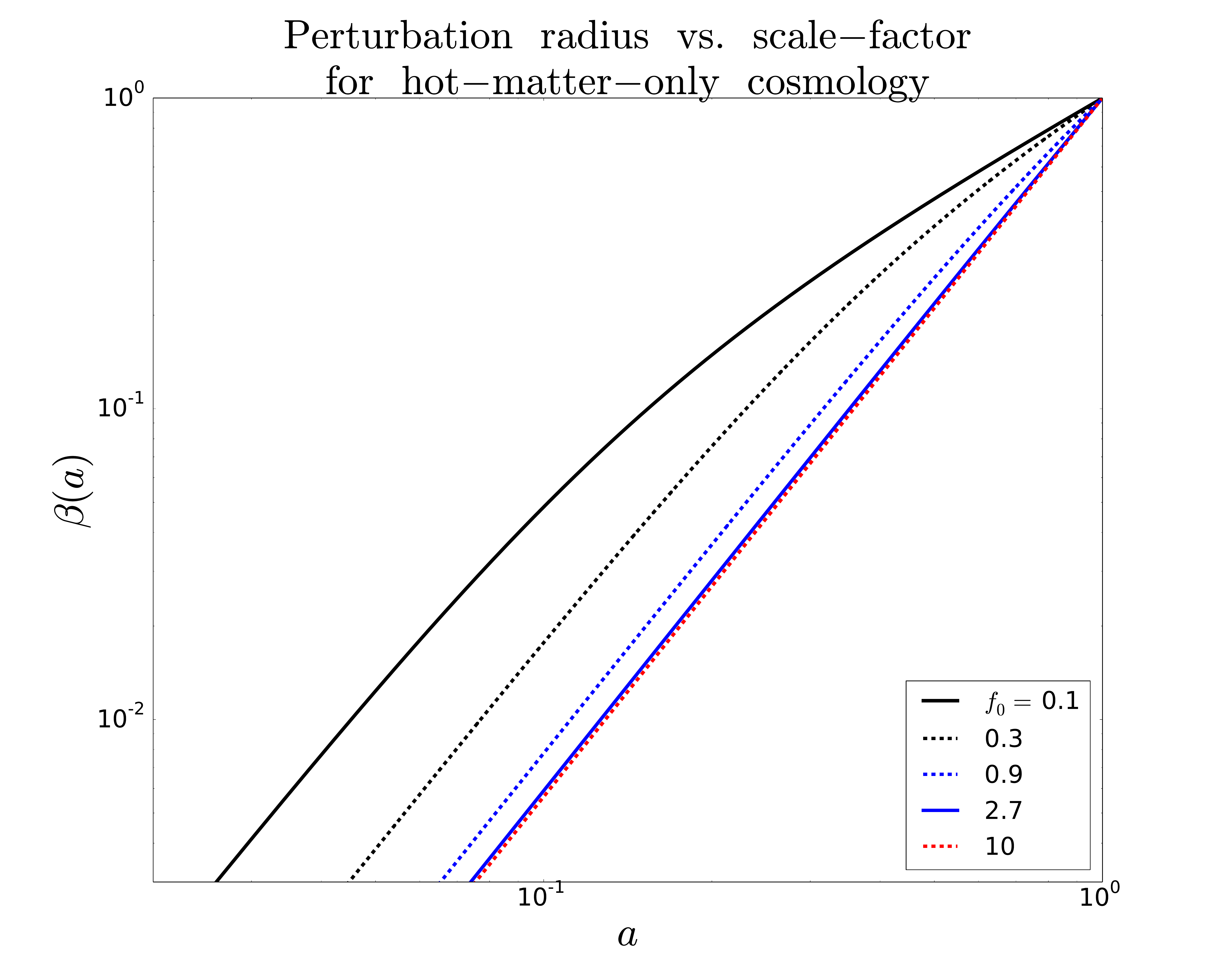}
\caption{Perturbation radius vs. scale-factor for illustrative values of $f_0$ (given in the legend) in the hot-matter-only cosmology. We have set $\beta_0 = 1$, where $\beta_0$ is the amplitude of the radial perturbation to the scale factor of our bubble universe at $a=1$, and again $\Omega_{{\rm X}0} = 1$. Since all perturbations must reach unity at present $(a=1)$, those with smaller amplitude at earlier times grow faster. We see that the largest $f_0$, more comparable to a radiation-only model in which perturbations would grow as $a^2$, grows faster than the smallest $f_0$ model, which is more comparable to a matter-only model in which perturbations grow as $a$. At early times, even the lowest-$f_0$ model behaves like radiation, shown by the fact that the black and dashed red curves have the same slopes for $a\lesssim 5\times 10^{-2}$. The two curves with the largest $f_0$ look very similar to each other because they have not yet become non-relativistic by the present and so both behave essentially as radiation.}
\label{fig:pert_hm_only}
\end{figure}

As $f_0\to \infty$ we should have the growth of perturbations in a radiation-only cosmology, i.e. $\beta(a)\propto a^2$ (see e.g. \cite{Padmanabhan}, equations 4.123 and 4.126).  We Taylor-expand $I_1(a)$ in the small parameter $\epsilon =a/f_0$, finding $I_1(a)\to a^4/8$ as $f_0\to\infty$.  Inserting this result in equation (\ref{eqn:perts_gen}) we recover the desired scaling.

We show the growth of $\beta$ with scale factor in Figure \ref{fig:pert_hm_only}. We choose a smaller range of values for $f_0$ here simply to facilitate comparison with the perturbation growth plots in the remainder of the paper, which use a smaller range of $f_0$ for reasons discussed at the end of \S\ref{sec:single_cmpt}. Figure \ref{fig:pert_hm_only} shows that the two cases that are still relativistic at present ($f_0 = 2.7$ and $f_0 = 10$) have essentially indistinguishable behavior, and grow as a power law $\beta \propto a^2$ as expected for growth of perturbations in a radiation dominated model. The smallest value of $f_0$ shown, $f_0 = 0.1$, is relativistic up to $a = f_0 = 0.1$ and grows as a radiation-like power law up to that time. They then begin to grow more slowly, as a matter-like power law $\beta \propto a$. This curvature of the growth history is clearly shown in the Figure. Finally, the other values of $f_0$ shown smoothly interpolate between the limiting cases of $f_0 = 10$ and $f_0 = 0.1$.

\section{Friedmann equation for a hot matter and radiation model}
\label{sec:hm_r}
We can generalize the model presented above to include a radiation component as well. Parametrizing the ratio of hot matter to radiation at present as $\mathcal{R}_{\rm x,0}$ the Friedmann equation becomes
\begin{align}
H^2=\left(\frac{1}{a}\frac{da}{dt}\right)^2 = H_0^2\omegar\left(\mathcal{R}_{\rm x,0}g^{1/2}(a)a^{-3} +a^{-4} \right).
\end{align}
$g(a)$ is as defined in equation (\ref{eqn:g_defn}). Taking the square root of both sides, rearranging, and integrating, we find for the time
\begin{align}
\frac{1}{\sqrt{\omegar}}\int_0^a \frac{a' da'}{\sqrt{\mathcal{R}_{\rm x,0}g^{1/2}(a')a'+1}}=I_{0,\rm{Xr}}(a)=H_0t.
\end{align}
The integral may be evaluated by defining the reduced ratio $\mu = \mathcal{R}_{\rm x,0}/\sqrt{1+f_0^2}$, making the change of variable $x=\sqrt{a'^2+f_0^2}$, and then making the further change of variable $u=\mu x+1$. Integrating we find
\begin{align}
I_{0,\rm Xr} = H_0t(a) = \frac{2}{\mu^2\sqrt{\omegar}}\left[\frac{1}{3}u^{3/2} -  u^{1/2}\right]\bigg|_{\mu f_0 + 1}^{\mu \sqrt{a^2 + f_0^2}+1}.
\label{eqn:Ixr}
\end{align}
In the limit that $f_0\to\infty$, we may Taylor expand in $\epsilon = a/f_0$; $\omegar \to1$ and $\mathcal{R}_{\rm x,0} \to 1$. The leading order series has $a^2\propto t$ and including the constants recovers the radiation-only solution.  

The $f_0\to0$ limit is more subtle; one then has cold, pressureless matter as well as radiation.  In this limit one recovers the scale-factor-time relation for such a cosmology, given in e.g. \cite{SEBAO} equation (14).  Note that \cite{SEBAO} sets the scale-factor to unity at matter-radiation equality, so to compare the limit here to that result one should set $\mathcal{R}_{\rm x,0}=1$, as is the case if $t_0$ is set to be matter-radiation equality.

\begin{figure}
\includegraphics[width=1.1\columnwidth]{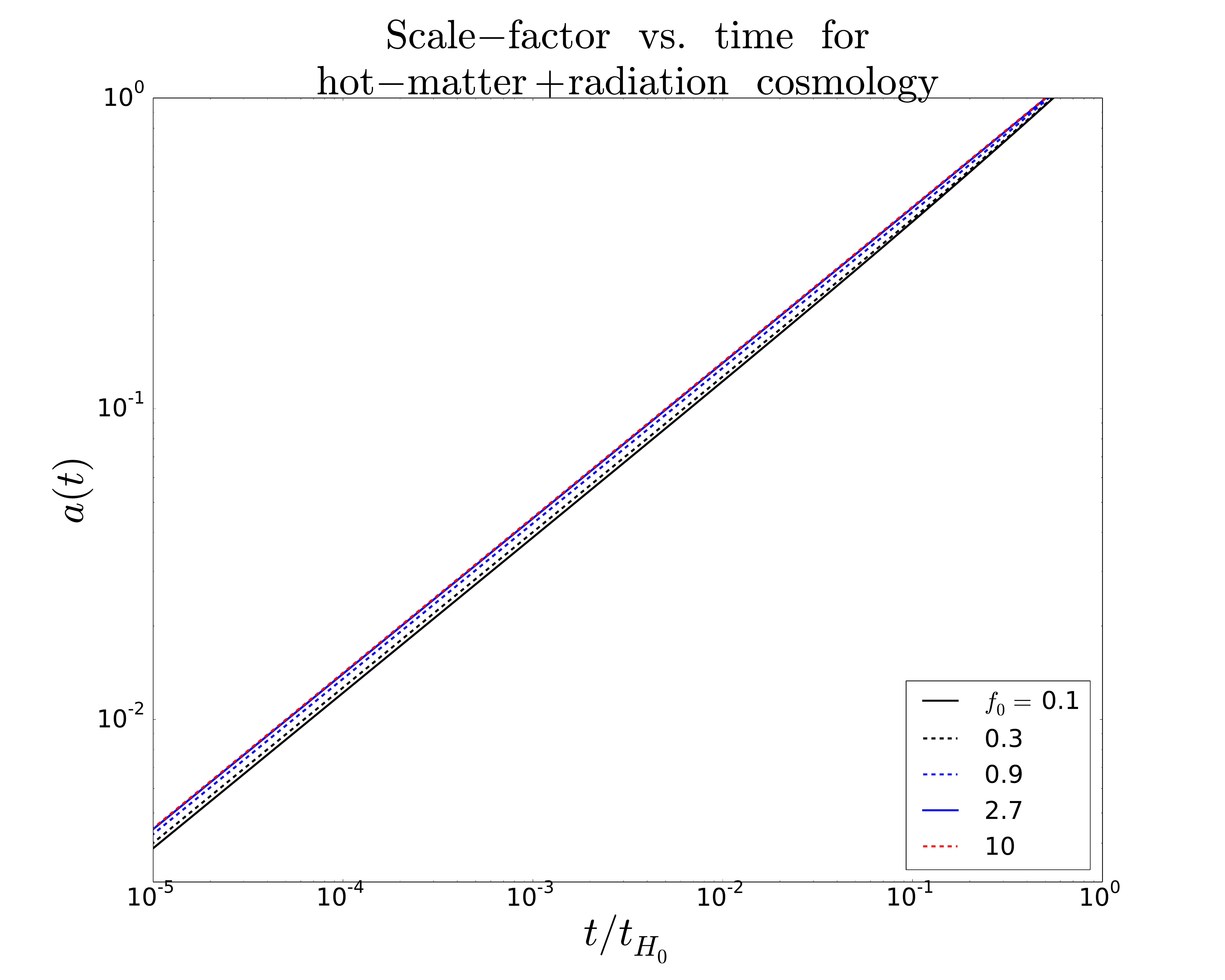}
\caption{Scale-factor vs. time for illustrative values of $f_0$ in the hot-matter plus radiation cosmology. We set $\Omega_{\rm X0} = 0.5$. For the largest value of $f_0$, the hot matter essentially always behaves as radiation and so the scale-factor always grows as the expected power law $a\propto t^{1/2}$, but for $f_0=0.1$, at late times the hot matter behaves as matter, steepening the power-law index of the scale factor from $1/2$ nearer to the $2/3$ expected for matter.}
\label{fig:hm_plus_r}
\end{figure}

Figure \ref{fig:hm_plus_r} shows the scale factor in this model.  The largest two values of $f_0$ are both relativistic even at present, and so the hot matter acts like radiation for the whole range of times displayed, leading to a power-law growth where $a$ scales roughly as $t^{1/2}$.  In the model with the smallest $f_0$, $f_0=0.1$, plotted in black, the curvature away from a power-law behavior at $a=f_0 = 0.1$ is evident. The other values of $f_0$ smoothly interpolate between these limiting cases. We note that $t_{H_0}$ is the same for all values of $f_0$ displayed, as it depends only on $\omegax$.  

\section{Growth of super-horizon perturbations in the hot matter-radiation model}
\label{sec:hm_r_perts}
The calculation for the growth rate of super-horizon perturbations in this cosmology proceeds analogously to that presented in \S\ref{sec:sh_perts_one_cmpt}.  Here, the analog of $I_0$ of equation (\ref{eqn:I0_I1_defn}) is $I_{0,\rm Xr}$, while the analog of $I_1$, which we denote $I_{1,{\rm Xr}}$, may be obtained from Taylor expanding a Friedmann equation with a curvature perturbation $C$.  The full, exact integral equation is
\begin{align}
\frac{1}{\sqrt{\omegar}} \int_0^r \frac{r' dr'}{\left[\mathcal{R}_{\rm x,0}g^{1/2}(r')r' + 1 + \tilde{C}r'^2\right]^{1/2}}=H_0t
\end{align}
where $\tilde{C} = C/\omegar$ is the curvature perturbation $C$ normalized by the radiation density parameter at present.  We now Taylor expand to leading order in $\tilde{C}$, finding 
\begin{align}
& I_{0,{\rm Xr}}(r) -\frac{\tilde{C}}{{2\sqrt{\omegar}}} \int_0^r \frac {r'^3dr'}{\left[ \mu \sqrt{r'^2 + f_0^2}+1\right]^{3/2}} \nonumber\\
& = I_{0,{\rm Xr}}(r) -\tilde{C}I_{1,{\rm Xr}}(r) = H_0t.
\end{align}
Using the same substitutions outlined for $I_{0,{\rm Xr}}$ (but with $a'\to r'$) the integral $I_{1,\rm Xr}$ becomes
\begin{align}
&I_{1,\rm Xr}(r) = \frac{1}{5\sqrt{\omegar}\mu^4 \sqrt{u}} \nonumber\\
&\times\bigg[5-5u^2+u^3-5f_0^2\mu^2 - 5u(f_0^2\mu^2-3)\bigg]\bigg|_{\mu f_0+1}^{\mu\sqrt{r^2+f_0^2}+1}.
\end{align}
Inserting the integral above as well as equation (\ref{eqn:Ixr}) for $I_{0,\rm Xr}$ into the first line of equation (\ref{eqn:perts_gen}) gives the growth of super-horizon perturbations in this model. Taking the limit as $f_0\to 0$ recovers the matter and radiation solution of \cite{SEBAO} equation (14) and taking the limit as $f_0 \to \infty$ gives $I_1\propto a^4$, leading to $\beta \propto a^2$ as expected for a radiation-only model.
\begin{figure}
\includegraphics[width=1.1\columnwidth]{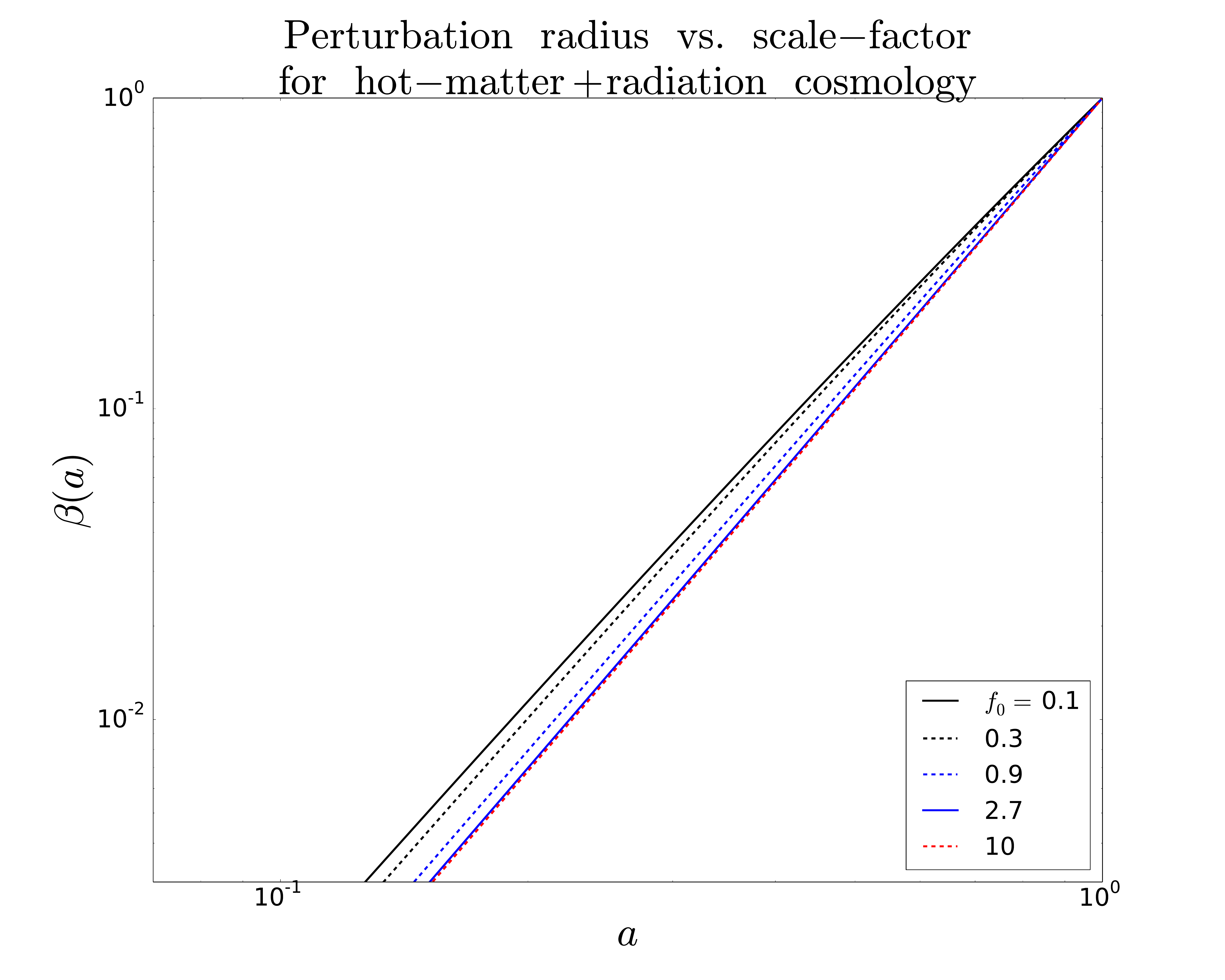}
\caption{Perturbation radius vs. scale-factor for illustrative values of $f_0$ in the hot-matter plus radiation cosmology. We have set the amplitude of the radial perturbation to the scale factor of the overdense bubble universe at $a=1$ as $\beta_0 = 1$. Similar to Figure \ref{fig:pert_hm_only}, we see that perturbations grow fastest in the model with largest $f_0$, and slowest in the model with smallest $f_0$, as expected from the radiation-only and matter-only limits.}
\label{fig:pert_hm_plus_r}
\end{figure}

Figure \ref{fig:pert_hm_plus_r} shows the growth of perturbations in the hot matter pus radiation model. The largest two $f_0$ values are relativistic even at present, and so the perturbation grows as $a^2$ as is expected for a radiation perturbation in a radiation-dominated model. This is also the behavior for $f_0 = 1$, but then at $a=f_0$ the growth switches over to that of a matter perturbation in a matter model, as $a$. This point also explains why the black curve is above the red $(f_0 = 10)$ curve at all times. We have normalized all perturbations to have $\beta=1$ at present, and so those growing more slowly (i.e. the $f_0 = 0.1$ model) had to begin at larger amplitude to reach unity at present.

\section{Friedmann equation for a hot matter and matter model}
\label{sec:hm_m}
The Friedmann equation for a hot matter and matter model is
\begin{align}
H^2 = \left(\frac{1}{a}\frac{da}{dt} \right)^2 = H_0^2 \omegam\left(\mathcal{R}_{\rm x, 0}g^{1/2}(a)a^{-3}+a^{-3}\right)
\end{align}
where now $\mathcal{R}_{\rm x,0}$ is the ratio of the hot matter to the matter density at $a=1$. Rearranging differentials, multiplying numerator and denominator by $a^{1/2}$, and using $\mu$ as defined in \S\ref{sec:hm_r}, we find
\begin{align}
I_{\rm 0Xm}(a)=\frac{1}{\sqrt{\omegam}}\int_0^a \frac{a'^{1/2}da'}{\sqrt{\mu\sqrt{1+f_0^2 a^{-2}}+1}}=H_0t.
\end{align}
Using the substitution $x=f_0^2 a'^{-2}$, so that $a'^{1/2}da' = -(1/2)f_0^{3/2}x^{-7/4}dx$, we obtain
\begin{align}
I_{\rm 0Xm}(a) = \frac{f_0^{3/2}}{2\sqrt{\omegam}}\int_{f_0^2a^{-2}}^{\infty}\frac{x^{-7/4}dx}{\sqrt{\mu\sqrt{1+x}+1}}.
\label{eqn:thirtyone}
\end{align}
Performing the integral and then defining the auxiliary variable $u=\sqrt{1+x}$ to shorten the result, we find
\begin{align}
&I_{\rm 0Xm}(a) = -\frac{2 f_0^{3/2}\sqrt{1+u\mu}}{3\sqrt{\omegam}(u^2-1)^{3/4}(\mu-1)^2(\mu + 1)}\nonumber\\
&\times \bigg\{ (\mu-1)(\mu u -1)+2^{1/4}\mu(u+1) \bigg[\frac{(u-1)(\mu-1)}{1+\mu u} \bigg]^{3/4} \nonumber\\
&\times  {_2F_1}\left(\frac{1}{4},\frac{3}{4},\frac{5}{4},\frac{(1+\mu)(1+u)}{2(1 + u\mu)}\right)\bigg\}\bigg|_{\sqrt{1+f_0^2 a^{-2}}}^{\infty},
\label{eqn:HM_M_integral}
\end{align}
where ${_2F_1}$ is Gauss's hypergeometric function and the sub- and superscripted bar gives the $u$ at which to evaluate the expression. We notice that the expression involves $(\mu -1)^{3/4}$ and that $\mu$ may be less than unity, leading to an imaginary pre-factor. We therefore apply Euler's third transformation\footnote{\url{http://mathworld.wolfram.com/EulersHypergeometricTransformations.html}, equation (8).} (\citealt{GR}, \citealt{WolframHGtrans}) to find 
\begin{align}
&I_{\rm 0Xm}(a) = -\frac{2f_0^{3/2}\sqrt{1+\mu u}}{3\sqrt{\omegam}(u^2-1)^{3/4}(\mu-1)^2(\mu+1)}\nonumber\\
&\bigg\{(\mu-1)(\mu u -1) + \mu(u+1)\left[ \frac{(u-1)(\mu-1)}{1+\mu u}\right]\nonumber\\
&\times _2F_1\left(1,\frac{1}{2},\frac{5}{4},\frac{(1+u)(1+\mu)}{2(1+\mu u)}\right)\bigg\}\bigg|_{\sqrt{1+f_0^2 a^{-2}}}^{\infty}
\label{eqn:transformed_HG_HM_plus_M}
\end{align}
We now check a few limits of this result to ensure the hypergeometric converges. As $\mathcal{R}_{\rm x,0}\to 0$ or $f_0\to \infty$, which are equivalent because both mean that there is only radiation in this model and no hot matter, $\mu\to 0$ and the hypergeometric's argument becomes $1/2$, for which it converges, as the hypergeometric has singular points only at $0,1$  and $\infty$. As $\mathcal{R}_{\rm x, 0}\to \infty$, i.e. there is only hot matter and no radiation in the model, $\mu \to \infty$, and the hypergeometric's argument becomes $1/2+ 1/(2u)$, which can reach unity if $u\to 1$. However, this occurs only as $a\to \infty$, so for finite scale factor we do not reach the singular point. Further, $u\to 1$ corresponds to $x\to 0$, and since the integrand of equation (\ref{eqn:thirtyone}) scales as $x^{-7/4}$ in this limit, we expect the integral to diverge. 

Indeed, $x\to 0$ may occur in two different ways: because $f_0^2 \to 0$ (the limit that the ``hot'' matter is actually cold matter), or because $a\to \infty$. From direct analysis of the integral in equation (\ref{eqn:thirtyone}), the divergence as $a\to \infty$ is a genuine one, whereas the $f_0^2 \to 0$ one is made finite by the $f_0^{3/2}$ pre-factor of equation (\ref{eqn:thirtyone}). In particular, making the substitution $f_0 = a\sqrt{x}$ and taking the limit of this equation gives $-2a^{3/2}/[3\sqrt{\omegam}\sqrt{1+\mu}]$ which is finite.

Finally, as $\mu \to 1$, the hypergeometric's argument goes to unity and it diverges; there is also a factor of $1/(\mu-1)^2$ in the pre-factor of equation (\ref{eqn:transformed_HG_HM_plus_M})  that is relevant in this limit. Examining our original equation (\ref{eqn:thirtyone}), we see no divergence as $\mu \to 1$, so we expect these divergences to cancel in the definite integral. This can be shown explicitly by taking a series for equation (\ref{eqn:transformed_HG_HM_plus_M}) about $\mu =1$. We find divergent terms proportional to $(\mu-1)^{-5/4}$ and $(\mu-1)^{-1/4}$, but these terms are independent of $u$ so when evaluated at the upper and lower bounds in equation (\ref{eqn:transformed_HG_HM_plus_M}) they cancel out, rendering our result finite. 

We also note that for $\mu=1$, the integral can in fact be done in elementary form as 
\begin{align}
I_{0 \rm Xm}(a) = \frac{4f_0^{3/2}}{30\sqrt \omegam}\left\{\frac{2f_0^2 +a\left(4a+\sqrt{a^2+f_0^2}\right)}{\sqrt{f_0^3\left(a+\sqrt{a^2+f_0^2}\right)}}-2 \right\}
\end{align}
and we suggest using this simpler result if $\mu=1$.\footnote{If evaluated in {\sc Mathematica} the more general form equation (\ref{eqn:transformed_HG_HM_plus_M}) will still give the correct answer up to a numerical-error-sized imaginary part that should be dropped.}

\begin{figure}
\includegraphics[width=1.1\columnwidth]{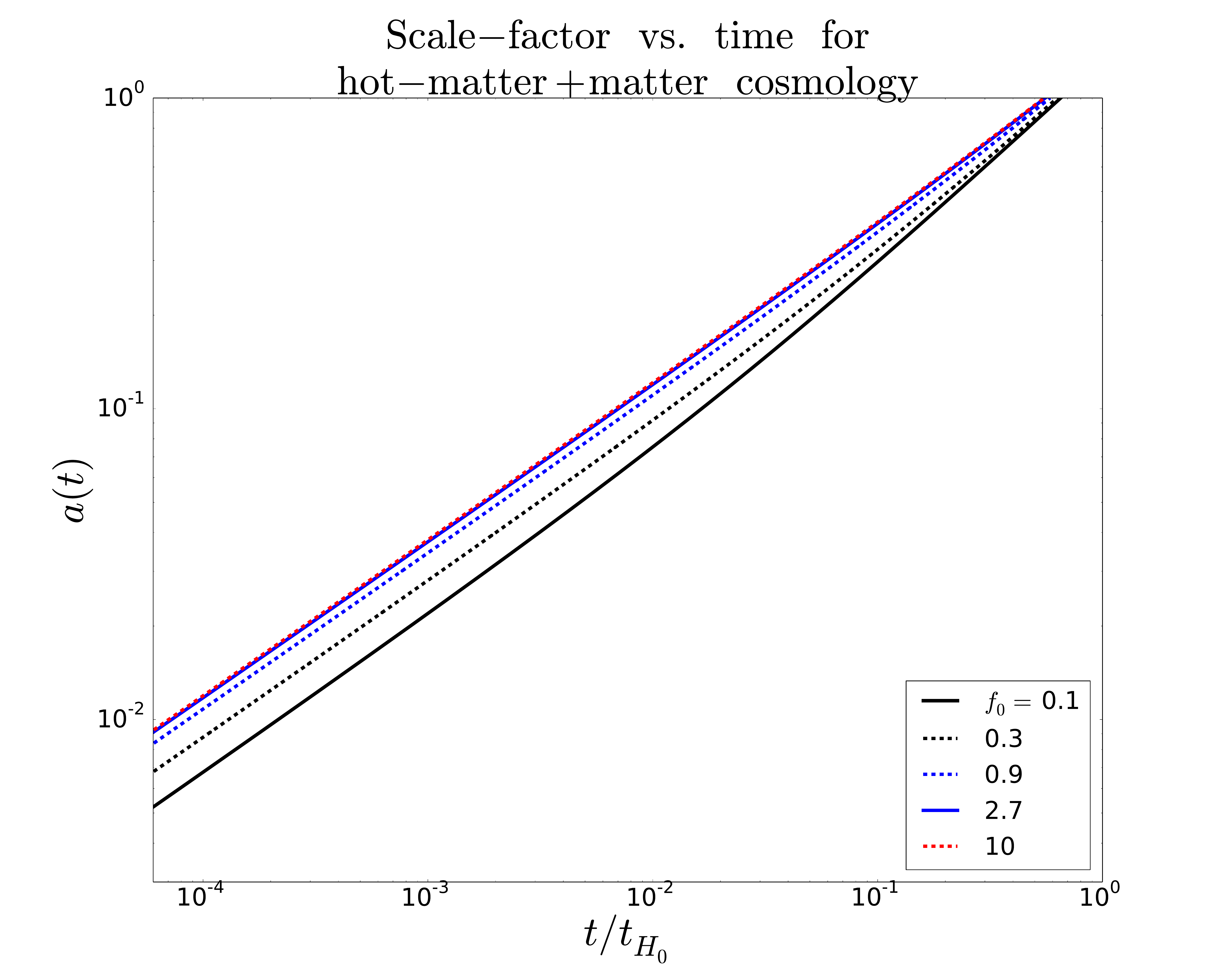}
\caption{Scale-factor vs. time for illustrative values of $f_0$ in the hot-matter plus matter cosmology. We set $\Omega_{\rm X0} = 0.5$. We see that for the lowest value of $f_0$ displayed, $f_0 =0.1$, the hot matter behaves as additional matter at late times and thus enhances the scale factor's growth, steepening the slope of the black curve relative to the others.}
\label{fig:hm_plus_m}
\end{figure}

Figure \ref{fig:hm_plus_m} shows the behavior of the scale factor in hot matter plus matter models. As in the previous figures, the smallest $f_0$ case behaves like radiation at early times and like matter at late times. The transition is shown in the curvature of the scale factor at $a = f_0 = 0.1$. As in the previous figures for the scale factor, the curves with go from smallest to largest $f_0$ as one goes from bottom to top. This is because  the scale factors are all normalized to reach unity at present, and models with a larger matter-like component grow more quickly at late times, and so can start with smaller $a$ and still reach $a=1$ today.

\section{Growth of super-horizon perturbations in the hot matter-matter model}
\label{sec:hm_m_perts}
The full equation for an overdense bubble universe with curvature perturbation $C$ is 
\begin{align}
I_{\rm Xm}(r) = \frac{1}{\sqrt{\omegam}}\int_0^r \frac{r'^{1/2}dr'}{\sqrt{\mathcal{R}_{\rm x,0}g^{1/2}(r')+1+\tilde{C}r'}} = H_0t,
\end{align}
where $\tilde{C}$ is the curvature perturbation normalized by the present-day matter density.
Taylor-expanding in $\tilde{C}$, we have
\begin{align}
&I_{\rm Xm}(r)\approx I_{0, \rm Xm}(r) -\frac{C}{2\sqrt{\omegam}}\int_0^r \frac{r'^{3/2}dr'} {\left[ \mathcal{R}_{\rm x,0}g^{1/2}(r') +1\right]^{3/2}}\nonumber\\
&=I_{0, \rm Xm}(r) - \tilde{C}I_{1, \rm Xm}(r).
\end{align}
We obtained $I_{0, \rm Xm}(r)$ in \S\ref{sec:hm_m}; we now obtain the integral proportional to $\tilde{C}$, which we denote $I_{1, \rm Xm}(r)$. Using the same substitutions as in \S\ref{sec:hm_m} we find
\begin{align}
I_{1, \rm Xm}(r) = \frac{f_0^{5/2}}{4\sqrt{\omegam}}\int_{f_0^2r^{-2}}^{\infty}\frac{x^{-9/4}dx}{\left[\mu\sqrt{1+x}+1 \right]^{3/2}}.
\end{align}
This integral can be performed by decomposition into partial fractions. Defining the auxiliary function $u(x)=\sqrt{1+x}$ (parallel to what was done in \S\ref{sec:hm_m}) to make the result more compact but also retaining $x$ where appropriate, we find
\begin{align}
&I_{1, \rm Xm}(r)=\frac{f_0^{5/2}(1+\mu)}{5x^{5/4}\sqrt{\omegam}}
\bigg[(1-\mu)\bigg(-u(x)+\mu(x+1)(4x+1)\nonumber\\
&+ \mu^2(1-7x)u(x) 
+\mu^3(x+1)(4x-1)\bigg) - 2^{3/4} \mu x(1+x+u(x))\nonumber\\
&\times (1+6\mu^2)\left[\frac{(u(x)-1)(\mu -1)}{1+\mu u(x)}\right]^{1/4}  \;_2F_1\left(\frac{1}{4},\frac{3}{4},\frac{7}{4},\frac{(1+u(x))(1+\mu)}{2(1+\mu u(x))} \right)\bigg]\nonumber\\
&\times\left[\sqrt{(1+x)(1+\mu u(x))}(1-\mu^2)^3 \right]^{-1}\bigg|_{f_0r^{-2}}^{\infty},
\end{align}
where the sub- and superscripted ending bar above means evaluation at the values of $x$ indicated. Like equation (\ref{eqn:HM_M_integral}), this result also becomes imaginary if $\mu <1$ due to the factor of $(\mu -1)^{1/4}$, but this can again be cured using the transformation made there. We obtain
\begin{align}
&I_{1, \rm Xm}(r)=\frac{f_0^{5/2}(1+\mu)}{5x^{5/4}\sqrt{\omegam}}
\bigg[(1-\mu)\bigg(-u(x)+\mu(x+1)(4x+1)\nonumber\\
&+ \mu^2(1-7x)u(x) 
+\mu^3(x+1)(4x-1)\bigg) -  \mu x(1+x+u(x))\nonumber\\
&\times(1+6\mu^2)\left[\frac{(u(x)-1)(\mu -1)}{1+\mu u(x)}\right]\; _2F_1\left(\frac{3}{2},1,\frac{7}{4},\frac{(1+u(x))(1+\mu)}{2(1+\mu u(x))} \right)\bigg]\nonumber\\
&\times\left[\sqrt{(1+x)(1+\mu u(x))}(1-\mu^2)^3 \right]^{-1}\bigg|_{f_0r^{-2}}^{\infty}.
\label{eqn:Ionexm_transf}
\end{align}
Analysis of the divergences here proceeds analogously to that for equation (\ref{eqn:transformed_HG_HM_plus_M}) since the argument of the hypergeometric is the same here. The behavior at $\mu =1$ again requires careful analysis, as inspection of the original integral indicates that there should not be a divergence. This can be shown explicitly by taking a series for equation (\ref{eqn:Ionexm_transf}) about $\mu =1$. We find divergent terms proportional to $(\mu-1)^{-11/4}$, $(\mu-1)^{-7/4}$, and $(\mu-1)^{-3/4}$, but these terms are independent of $x$ so when evaluated at the upper and lower bounds in equation (\ref{eqn:Ionexm_transf}) they cancel out, rendering our result finite. 

Indeed, a simple elementary form is available at $\mu =1$, as
\begin{align}
&I_{1, \rm Xm}(r) = \frac{f_0^{5/2}}{4\sqrt{\omegam}}
\bigg\{4 \bigg(7 (8 + 3 u) - 6 ( u^2-1) \big[1 + 12 u + 8 u^2 \nonumber\\
&- 8 (1 + u)^{3/2} (  u^2-1)^{1/4}\big]\bigg)\bigg\}\bigg[385 (1 + u)^{3/2} ( u^2-1)^{5/4}\bigg]^{-1}
\end{align}
here with $u = \sqrt{1+f_0 r^{-2}}$.
\begin{figure}
\includegraphics[width=1.1\columnwidth]{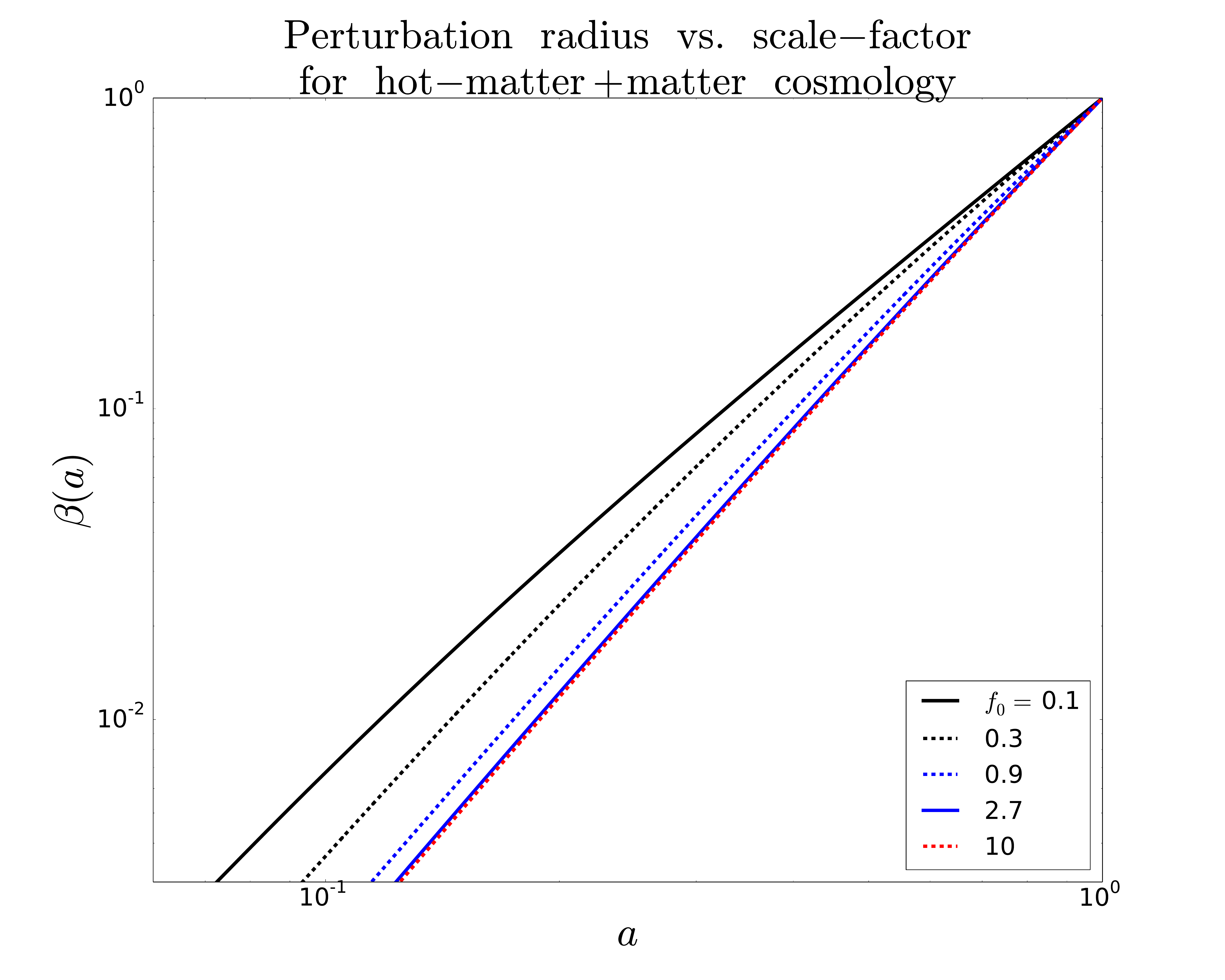}
\caption{Perturbation radius vs. scale-factor for illustrative values of $f_0$ in the hot-matter plus matter cosmology. We have set the amplitude of the radial perturbation to the overdense bubble universe's scale factor as $\beta_0 = 1$ at $a=1$. Similar to Figure \ref{fig:pert_hm_only}, we see that perturbations grow fastest in the model with largest $f_0$, and slowest in the model with smallest $f_0$, as expected from the matter plus radiation and matter-only limits.}
\label{fig:pert_hm_plus_m}
\end{figure}

Figure \ref{fig:pert_hm_plus_m} shows the growth of perturbations in a hot matter plus matter model. Again the smallest value of $f_0$, shown in black, displays the most interesting behavior, transitioning from the $\beta \propto a^2$ growth of radiation perturbations in a radiation-dominated model to the slower $\beta \propto a$ growth of perturbations in a matter-dominated model.

\section{Generalizing to more realistic momentum distributions}
\label{sec:approx_conv}
Up to this point, we have considered a species whose distribution of momentum magnitudes is a Delta function at $p_0 =f_0 mc$. We now seek to relax this assumption and generalize our work to more realistic momentum distributions, such as Fermi-Dirac (for neutrinos) or Bose-Einstein.  

The difficulty of doing so stems from the non-linearity of the Friedmann equation in the densities. Were the equation linear, one could solve for a Delta-function momentum distribution and then convolve with the true distribution. However, the non-linearity of the equation renders this impossible.

In detail, to solve the Friedmann equation one integrates in $a$, and properly handling an arbitrary momentum distribution would promote $g^{1/2}(a)$ to an integral of $g^{1/2}(a)$ against the distribution function. This integral would then appear inside a square root in the denominator of the further integral over $a$ required for finding $t(a)$, i.e solving the Friedmann equation by quadrature. Consequently one cannot interchange the order of integration to handle the distribution function after solving for $t(a)$. Rather one must integrate over the distribution function first. However, this would lead to Fermi-Dirac or Bose-Einstein integrals with $a$-dependence, which would be intractable to then integrate over $a$ to obtain $t(a)$. 

Explicitly, the energy density of the hot matter is given by a generalization of equation (\ref{eqn:rho_of_a}),
\begin{align}
\rho(a)&=\int_0^{\infty} \frac{d\rho_0}{df_0}g^{1/2}(a)a^{-3} df_0\nonumber\\
&=\int_0^{\infty} \frac{dn_0}{df_0}mc^2\sqrt{1+f_0^2}g^{1/2}(a)a^{-3} df_0\nonumber\\
&=n_0mc^2a^{-3}\int_0^{\infty}\mathcal{F}(f_0)\sqrt{1+f_0^2a^{-2}}df_0,
\label{eqn:rho_of_a_full}
\end{align} 
with $\mathcal{F}(f_0)$ being the distribution function in terms of $f_0 = p_0/(mc)$, normalized such that $\int_0^{\infty} \mathcal{F}(f_0) df_0 = 1$. 

In this notation, the density of hot matter at $a=1$ is
\begin{align}
\rho_{\rm 0} = n_0mc^2\int_0^{\infty}\mathcal{F}(f_0)\sqrt{1+f_0^2}df_0 \equiv n_0 mc^2 N_D.
\label{eqn:ND}
\end{align}
We note that in the limit where $\mathcal{F}$ is a Delta-function about $f_0 =0$, i.e. the particles are all at rest, we recover the expected energy density $n_0mc^2$. $N_D$ can thus be interpreted as a correction that scales up the energy density to account for the particles' kinetic energy.

The Friedmann equation now becomes
\begin{align}
H_0t = \int_0^a \frac{da'}{a' \sqrt{\omegam a'^{-3} + \omegar a'^{-4}+\omegax a'^{-3}I_{\rm D}(a)}}
\label{eqn:formal_soln_fmann}
\end{align}
with 
\begin{align}
I_{\rm D}(a) &= \frac{1}{N_D} \int_0^{\infty} \mathcal{F}(f_0) \sqrt{1+f_0^2 a'^{-2}} df_0.
\label{eqn:bad_int}
\end{align}

Consider the case of the cosmic neutrino background. Since the neutrinos decouple while they are still relativistic, their momentum distribution at decoupling is a Fermi-Dirac distribution
\begin{align}
\mathcal{F}(p) dp = \frac{2 c^3}{3 \zeta(3) (\kB T_{\rm{dec}})^3} \frac{p^2 dp}{\exp\left[p c/(\kB T_{\rm{dec}})\right] + 1}
\end{align}
with $T_{\rm{dec}}$ the temperature of neutrino decoupling and $\kB$ is the Boltzmann constant. The chemical potential of the neutrinos before decoupling is $\mu=0$ assuming that they are in equilibrium with the radiation. Let the temperature of neutrinos at present be $T_{\rm{X0}}$, so the present momentum a neutrino with momentum $p$ at decoupling is $p_0 = p\left(T_{\rm{X0}}/T_{\rm{dec}}\right)$. Then the distribution function at present in terms of $f_0$ is
\begin{align}
\mathcal{F}(f_0) df_0 = \left(\frac{mc^2}{\kB T_{\rm{X0}}}\right)^3 \frac{2}{3 \zeta(3)} \frac{f_0^2 df_0}{\exp\left[f_0mc^2/(\kB T_{\rm{X0}})\right]+1}.
\label{eqn:fermi_dirac_dist}
\end{align}

\subsection{Choice of central momentum for Delta-function approximation}
We numerically calculate the difference in $t(a)$ obtained using this Fermi-Dirac distribution instead of a Delta-function momentum distribution. For the Delta-function momentum distribution, we want to find a momentum $\hat{f}_0$ that will approximate the Fermi-Dirac distribution well. We choose $\hat{f}_0$ such that the energy density calculated using the Delta-function distribution (equation (\ref{eqn:rho_of_a})) agrees with the energy density of the Fermi-Dirac distribution (equation (\ref{eqn:rho_of_a_full})) at early times. At late times $a >> f_0$, both distributions already agree, giving $\rho(a) \approx n_0 mc^2 a^{-3}$ as expected for cold matter. At early times $a << f_0$, the Delta-function gives
\begin{align}
\rho(a) \approx \frac{\rho_0 \hat{f}_0 a^{-4}}{1+\hat{f}_0^2} = n_0 mc^2 \hat{f}_0 a^{-4}
\end{align}
while the full distribution yields
\begin{align}
\rho(a) \approx n_0 mc^2 a^{-4} \int_0^{\infty} \mathcal{F}(f_0) f_0 df_0.
\end{align}
Setting these expressions for $\rho(a)$ equal, we find
\begin{align}
\hat{f}_0 = \int_0^{\infty} \mathcal{F}(f_0) f_0 df_0 = \frac{7 \pi^4}{180 \zeta(3)} \frac{\kB T_{\rm X0}}{mc^2} \approx 3.152 \frac{\kB T_{\rm X0}}{mc^2},
\label{eqn:fhat}
\end{align}
the average neutrino momentum today. 

\begin{figure}
\includegraphics[width=\columnwidth]{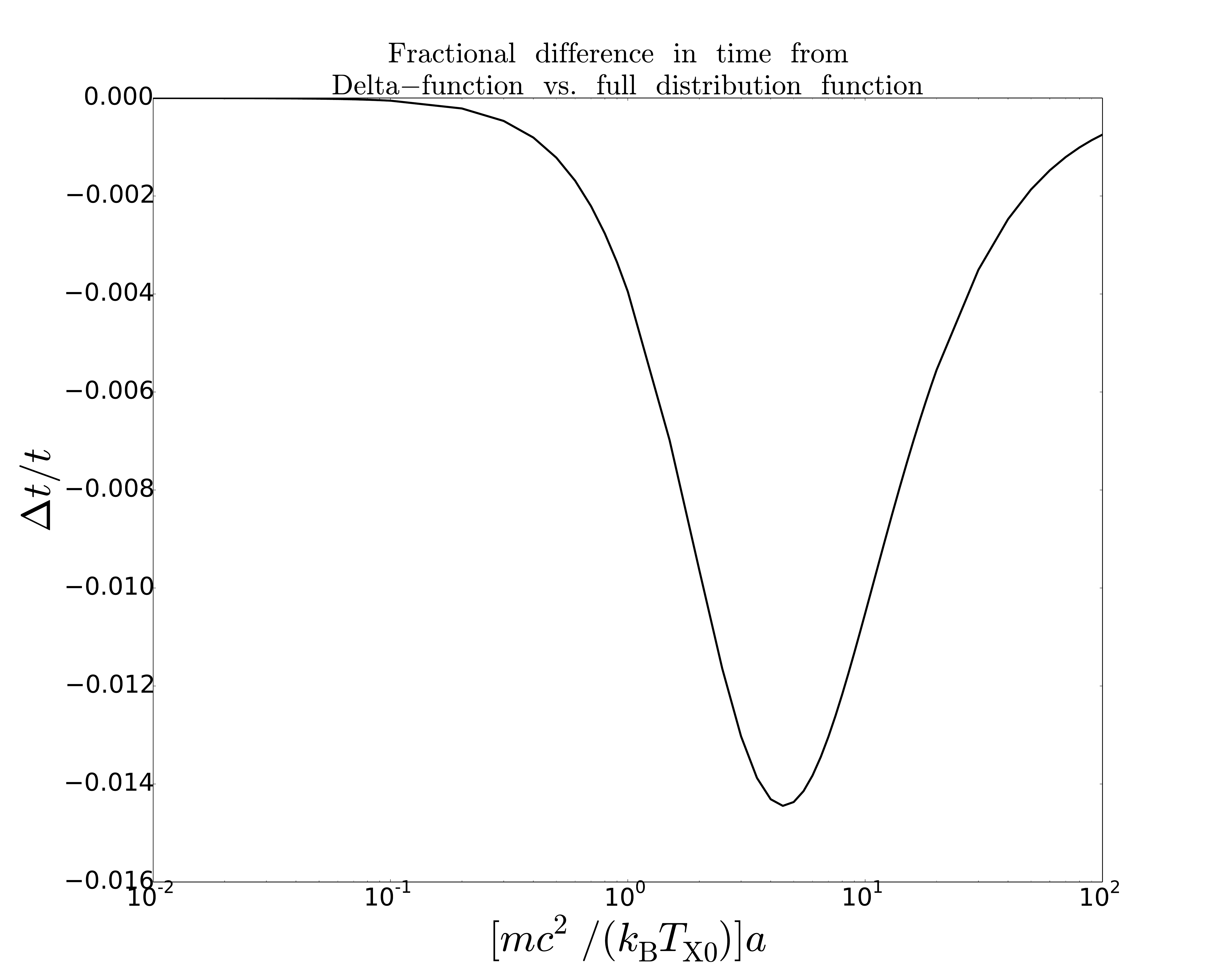}
\caption{Fractional difference in time as a function of scale factor using the Delta-function momentum distribution approximation versus the full Fermi-Dirac distribution. The fractional difference is at most $1.5\%$, and this occurs at the redshift when the hot matter transitions over from being relativistic and radiation-like to being cold and matter-like, when $a=\hat{f}_0 = 3.152 k_{\rm B}T_{\rm X0}/(mc^2)$ from equation (\ref{eqn:fhat}). We might expect this to be the point where there is the largest error in our approximation because the width of the distribution function will set the detailed behavior of this transition, and our Delta-function approximation does not contain any information about this width.}
\label{fig:fd_timediff}
\end{figure}

The fractional difference in $H_0 t(a)$ due to the neutrino momentum distribution will be greatest in a neutrino-only universe, reducing equation (\ref{eqn:formal_soln_fmann}) to
\begin{align}
H_0 t = \int_0^a \frac{a'^{1/2} da'}{\sqrt{I_{\rm D}(a)}},
\end{align} which for the case of the Fermi-Dirac distribution can be solved numerically. Comparing $t(a)$ for the Fermi-Dirac distribution and the Delta-function distribution with $\hat{f}_0$ given by equation (\ref{eqn:fhat}) in Figure \ref{fig:fd_timediff}, we find that the fractional difference never exceeds 1.5\%.

\subsection{Incorporating a finite width for the momentum distribution}
\label{subsec:width}
We can also estimate the effect of the distribution function generally, beyond specific functional forms like the Fermi-Dirac distribution.
If the momentum distribution is peaked about some momentum $\hat{f}_0$, then equation (\ref{eqn:rho_of_a_full}) can be Taylor-expanded about $f_0=\hat{f}_0$, yielding
\begin{align}
&\rho(a) \approx n_0 mc^2 a^{-3} \int_0^{\infty} \mathcal{F}(f_0) \bigg(\sqrt{1 + \hat{f}_0^2 a^{-2}}\nonumber\\
&+ \frac{\hat{f}_0 a^{-2}}{\sqrt{1+\hat{f}_0^2 a^{-2}}} (f_0 - \hat{f}_0)+ \frac{a^{-2}}{2(1+\hat{f}_0^2 a^{-2})^{3/2}} (f_0-\hat{f}_0)^2 \bigg) df_0
\end{align}
The first term is simply the energy density under a Delta-function momentum distribution at $\hat{f}_0$, which we will denote $\hat{\rho}(a)$. Using $\hat{f}_0$ as found in equation (\ref{eqn:fhat}), the second term is zero, so the energy density is
\begin{align}
\rho(a) &\approx \hat{\rho}(a) \left(1 + \frac{a^{-2}}{2(1+\hat{f}_0^2a^{-2})^{2}} \int_0^{\infty}\mathcal{F}(f_0) (f_0 - \hat{f}_0)^2 df_0 \right)\nonumber\\
 &= \hat{\rho}(a) \left(1 + \frac{\int_0^{\infty}\mathcal{F}(f_0) f_0^2 df_0 - \hat{f}_0^2}{2a^2(1+\hat{f}_0^2a^{-2})^{2}}\right)
\label{eqn:rho_frac_diff}
\end{align}
The fractional difference between $\rho(a)$ and $\hat{\rho}(a)$ goes to zero at early and late times. From finding the minimum of the denominator in equation (\ref{eqn:rho_frac_diff}) we see that the fractional difference is maximized at $a=\hat{f}_0$, yielding:
\begin{align}
\left.\frac{\rho(a)-\hat{\rho}(a)}{\hat{\rho}(a)}\right|_{a=\hat{f}_0} &\approx \frac{\int_0^{\infty}\mathcal{F}(f_0) f_0^2 df_0 - \hat{f}_0^2}{8 \hat{f}_0^2} \nonumber\\
&= \frac{1}{8}\left(\frac{\int_0^{\infty}\mathcal{F}(f_0) f_0^2 df_0}{\hat{f}_0^2}-1\right).
\label{eqn:max_frac_diff_rho}
\end{align}
This is $1/8$ the ratio of the variance to the square of the mean. This result holds for any distribution function with well-defined second moment; in particular it would hold for a Bose-Einstein distribution.

As an example, consider the Fermi-Dirac distribution which is appropriate for the cosmic neutrino background. Substituting this distribution (equation (\ref{eqn:fermi_dirac_dist})) and our chosen $\hat{f}_0$ (equation (\ref{eqn:fhat})) into equation (\ref{eqn:max_frac_diff_rho}) gives
\begin{align}
\left.\frac{\rho(a)-\hat{\rho}(a)}{\hat{\rho}(a)}\right|_{a=\hat{f}_0} \approx \frac{1}{8} \left[\frac{15\zeta(5)}{\zeta(3)}\left(\frac{180 \zeta(3)}{7 \pi^4}\right)^2 - 1\right]\approx 0.038.
\end{align}
For a Bose-Einstein distribution, the analogous result is $0.052$.
The Friedmann equation (\ref{eqn:formal_soln_fmann}) depends on the square root of the energy density, so taking a leading-order Taylor series the error due to dropping the second moment should be $(1/2)\times 0.038 = 1.9\%$. Thus our finding that the time $t(a)$ differs by at most 1.5\% is reasonable. For a Bose-Einstein distribution, we would expect a difference of at most approximately $(1/2)\times 0.052 = 2.3\%$.

If the full distribution function only produces a small fractional change to the energy density of the hot matter relative to the Delta-function result, the Friedmann equation can be linearized in terms of this fractional change. For example, consider a hot matter-only universe and add the fractional energy density change from equation (\ref{eqn:rho_frac_diff}):
\begin{align}
H_0 dt = \frac{da}{a} \left[a^{-3}g^{1/2}(a) \left(1 + \frac{\int_0^{\infty}\mathcal{F}(f_0) f_0^2 df_0 - \hat{f}_0^2}{2a^2(1+\hat{f}_0^2a^{-2})^{2}}\right)\right]^{-1/2}.
\end{align}
Taylor-expanding in the fractional energy density change yields
\begin{align}
H_0 dt &= \frac{a^{1/2} da}{g^{1/4}(a)} \left(1 - \frac{\int_0^{\infty}\mathcal{F}(f_0) f_0^2 df_0 - \hat{f}_0^2}{4a^2(1+\hat{f}_0^2a^{-2})^{2}}\right),
\end{align}
which can be integrated as
\begin{align}
H_0 t(a) &= I_0(a) + \frac{(1+\hat{f}_0^2)^{1/4}}{10} \nonumber\\ & \times \left(\frac{4}{\sqrt{\hat{f}_0}}-\frac{5a^2+4\hat{f}_0^2}{(a^2+\hat{f}_0^2)^{5/4}}\right)\left(\int_0^\infty \mathcal{F}(f_0) f_0^2 df_0 - \hat{f}_0^2\right).
\end{align}
Regardless of the exact form of the distribution function, the highest-order term in $H_0 t(a)$ arising from it has the above dependence on $a$ and $\hat{f}_0$.
\subsection{Exact solution for an arbitrary momentum distribution}
\label{subsec:exact}
We have shown that a second-order correction reflecting the momentum distribution's width can be incorporated and the Friedmann equation solved including this correction. We now generalize this idea to consider an expansion to all orders of the momentum distribution in terms of its moments.

We begin with the full energy density (\ref{eqn:rho_of_a_full}). Our strategy will be to Taylor-expand the function multiplying the distribution function $\mathcal{F}$ in the integrand; we define this function as
\begin{align}
h(f_0, a) \equiv \sqrt{1 + f_0^2 a^{-2}}.
\end{align}
We Taylor-expand $h$ in $f_0$ about $\hat{f}_0$; the powers of $f_0 - \hat{f}_0$ that result may then be integrated against the distribution function $\mathcal{F}$ to yield a series in moments of $\mathcal{F}$. Finally we divide out the contribution due to the zeroth moment of the distribution function $\hat{f}_0$; this contributions stems from the Dirac-Delta-function pieces. With these manipulations, the energy density (\ref{eqn:rho_of_a_full}) becomes
\begin{align}
\rho(a) = &\rho_{00}g^{1/2}(a)a^{-3}\bigg[1+g^{-1/2}(a)\nonumber\\
&\times\sum_{n=1}^{\infty}\left(\frac{\partial h}{\partial f_0}\right)^n\bigg|_{f_0 = \hat{f}_0} M_n[\mathcal{F}](\hat{f}_0) \bigg]
\label{eqn:rho_a_full}
\end{align}
where $\rho_{00}=n_0mc^2$ is the present-day energy density of the species due to the centroid of its momentum distribution and the $n^{th}$ moment of the distribution function is
\begin{align}
M_n[\mathcal{F}](\hat{f}_0) = \frac{1}{N_D}\int_0^{\infty}\mathcal{F}(f_0)(f_0-\hat{f}_0)^n df_0,
\end{align}
with $N_D = \rho_{0}/\rho_{00}$ by inserting the defintion of $\rho_{00}$ in equation (\ref{eqn:ND}). For a distribution symmetric about $\hat{f}_0$ only the even moments will be non-zero.

We now insert equation (\ref{eqn:rho_a_full}) for $\rho(a)$ in the Friedmann equation, rearrange, and Taylor expand taking it that the sum of the corrections due to the distribution function is small. We find
\begin{align}
H_0t &= I_0(a)-\frac{1}{2}\sum_{n=1}^{\infty}M_n[\mathcal{F}](\hat{f}_0)\int_0^{a}a'^{1/2}g^{-3/4}(a')\left(\frac{\partial h}{\partial f_0} \right)^n\bigg|_{f_0 = \hat{f}_0}(a)\nonumber\\
&\equiv I_0(a) -\frac{1}{2}\sum_{n=1}^{\infty}\left(\frac{\partial }{\partial f_0} \right)^n\bigg|_{f_0 = \hat{f}_0}I_1(a;f_0),
\label{eqn:soln_w_dist}
\end{align}
with 
\begin{align}
&I_1(a;f_0) = (1+\hat{f}_0^2)^{3/4}\int_0^a a'^{1/2}(1+f_0^2a'^{-2})^{-1/4}da'\nonumber\\
&=\frac{2}{3}(1+\hat{f}_0^2)^{3/4} \left[(f_0^2+a^2)^{3/4}-f_0^{3/2}\right].
\end{align}
We obtained the second line of equation (\ref{eqn:soln_w_dist}) by interchanging the derivative with respect to $f_0$ and the integration with respect to $a$ in the first line, so that the behavior of $H_0t$ sourced by each moment of the distribution function could be cast as derivatives of a single fundamental function $I_1(a; f_0)$. Our full solution to the Friedmann equation for arbitrary momentum distributions is thus
\begin{align}
&H_0t = I_0(a) -\frac{1}{3}(1+\hat{f}_0^2)^{3/4}\sum_{n=1}^{\infty}M_n[\mathcal{F}](\hat{f}_0)\nonumber\\
&\times\left(\frac{\partial}{\partial f_0} \right)^n\bigg|_{f_0 = \hat{f}_0}\left[(f_0^2+a^2)^{3/4}-f_0^{3/2}\right]
\end{align}
This solution is valid as long as the expansion of $\mathcal{F}$ in terms of its moments converges. We observe that the larger $n$, the more powers of scale factor enter with negative power, so that the importance of higher terms in the moment expansion is suppressed as $a$ grows at later times, as we might have expected since peculiar velocities redshift away.

\section{Conclusions \& Discussion}
We have shown that for a species with a Dirac-delta-function distribution of momentum magnitudes, the equation of state assumes a simple form that at high redshift acts as radiation and at low redshift acts as matter. This equation of state can be analytically integrated to yield the evolution of the specie's energy density with scale factor, and this can then be inserted into the Friedmann equation. For a number of cases, the Friedmann equation can then be analytically solved by quadrature for the  relation between scale factor and time. 

This toy model approximately describes both neutrinos and warm dark matter, each of which go from acting as relativistic species at high redshift to acting as cold, clustering species with equation of state near zero at present.  The primary outcome of this work is to show that the impact of these species on the cosmological expansion of the Universe can be simply understood via relatively compact closed-form solutions.  

With the increased attention that will come both to massive neutrinos, with upcoming surveys such as DESI targeted to determine their mass sum, and WDM, with ever-tightening constraints from cosmological probes such as the Lyman-$\alpha$ forest, we believe it is timely to have an analytic model for their effects. While the current constraints on these species are surely correctly evolving the Hubble rate numerically, an analytic form can provide valuable intuition and perhaps inspire novel additional probes. 

We show that for a cosmology with hot matter alone, the reconstruction of scale factor versus time for this Delta-function distribution toy model is highly accurate at all times, differing from the solution for themore realistic Fermi-Dirac distribution (for neutrinos) by less than $1.5\%$, and generally much less. The worst performance is at the redshift of the species' transition from relativistic to non-relativistic. Further, in more realistic models including standard dark matter, dark energy, and radiation, this deviation would be suppressed as the neutrinos will not be the dominant driver of the cosmological expansion so the contribution to the scale factor's evolution of any error in their treatment is greatly reduced.

An important additional facet of this work is the moment approach we introduce in the final section, showing that one can perturbatively solve the Friedmann equation including higher moments of the hot matter's distribution function, as long as these higher moments are small compared to the mean.  We do this explicitly for the second moment and then show how to extend the treatment for arbitrary moments. We suggest this as a technique for using expansion rate measurements to constrain DM models with arbitrary distribution functions (e.g. \citealt{Boyanovsky08} \citealt{Boyanovsky11}), or neutrinos with non-standard statistics (e.g. \citealt{Miranda15}). One might implement the solution to the Friedmann equation provided here but leaving each moment of the distribution free (or imposing any desired recursion on the moments) and embed this in a cosmological parameter Markov chain Monte Carlo.

A possible direction of future work is to extend the toy model presented here to three species of neutrinos with different masses. There are two qualitative cases, the first with two neutrinos being heavy and one light (inverted hierarchy), and the second with one neutrino being heavy and the other two light (normal hierarchy). We expect that our model as is would better describe the second of these cases, as the two light species would likely still be relativistic at present, or at least transition to being non-relativistic late enough that they have already become a very negligible part of the total energy density (as they would have evolved like radiation up to that point). 

In the first case, where two neutrinos are heavy, if the two heavy neutrinos have somewhat different masses, this would effectively blur the redshift of their transition from relativistic to non-relativistic. It might be modeled as having a single effective redshift but with a larger width to the transition than the pure Delta-function-momentum distribution toy model we have considered. Including the full distribution function of the neutrinos, as discussed in \S\ref{sec:approx_conv}, would also have this effect, as neutrinos with lower kinetic energy would transition from relativistic to non-relativistic earlier and with higher kinetic energy would transition later.  

A more mathematical way to phrase this comment is that the transition occurs when $f_0 = p_0c/(mc^2)$ crosses unity, and this crossing can have a width either due to the numerator $p_0$ having a width from the distribution function, or due to the denominator $m$ effectively having a width due to the presence of two different masses. 

This qualitative discussion indicates that to constrain the neutrino mass splitting via its effect on the cosmological expansion rate will require both precise observations and precise modeling to disentangle this effect from that of the distribution function.  

Finally, each of the neutrino mass species is $\sim$$22\%$ of the total energy density in photons during radiation domination, and this suggests an avenue for perturbatively treating the three masses. One might write the neutrino energy density as the sum of these three components, and then Taylor-expand the Friedmann equation in these three components as fractions of the total energy density in matter and radiation. At early times, these fractions are small, and including the matter guarantees that the fractions remain small, and the expansion valid, to late times. To solve the Friedmann equation by quadrature in this approximation, but still taking it that the distribution function is a Delta function, one has a correction to the no-neutrino solution as $-(1/2)\Omega_{\rm X,i} \int a^{-4}\sqrt{g(a)}da/[\Omega_{\rm m0}a^{-3} + \Omega_{\rm r0}a^{-4}]^{3/2}$ with $\Omega_{\rm X,i}$ the energy density in the $i^{th}$ neutrino species at present. Further study of this solution including the effects of a more realistic distribution function may be an avenue of future work, though full analysis of the very high-precision future surveys required for constraining the mass splitting will certainly demand an exact numerical treatment as well.

As massive neutrinos grow in importance as a research topic in cosmology, we hope the treatment presented here will be of use as a simplified yet still fairly accurate picture of their effects on the cosmological expansion rate and the growth of structure.  

\section*{Acknowledgements}
We thank Simeon Bird, Emanuele Castorina, Daniel J. Eisenstein, Douglas P. Finkbeiner, Dan Green, Charles-Antoine Collins-Fekete, Simone Ferraro, Martina Gerbino, Elena Giusarma, Chris Hirata, Hamsa Padmanabhan, Matthew Pasquini, Uro\v s Seljak, Masahiro Takada, Martin White, Sunny Vagnozzi, and Matias Zaldarriaga for useful discussions. Support for this work was provided by the National Aeronautics and Space Administration through Einstein Postdoctoral Fellowship Award Number PF7-180167 issued by the Chandra X-ray Observatory Center, which is operated by the Smithsonian Astrophysical Observatory for and on behalf of the National Aeronautics Space Administration under contract NAS8-03060. ZS also acknowledges support from a Chamberlain Fellowship at Lawrence Berkeley National Laboratory (held prior to the Einstein) and from the Berkeley Center for Cosmological Physics. SKNP is supported in part by a Natural Sciences and Engineering Research Council of Canada Postgraduate Scholarship.




\bibliographystyle{mnras}
\bibliography{hotmatter}




\bsp	
\label{lastpage}
\end{document}